\documentclass[twocolumn]{aastex62}
\usepackage{natbib}
\usepackage{hyperref}

\newcommand\clj{SPT0615$-$57}
\newcommand\longname{SPT-CLJ0615$-$5746}

\graphicspath{{./}{figures/}}

\received{2019 March 4}
\revised{2019 April 29}
\accepted{2019 May 1}
\submitjournal{The Astrophysical Journal}

\shorttitle{Kinematics of CLJ0615$-$57}
\shortauthors{Connor et al.}

\begin{document}

\title{Assembling a RELIC at Redshift 1: Spectroscopic Observations of Galaxies in the RELICS Cluster SPT-CLJ0615$-$5746}

\correspondingauthor{Thomas Connor}
\email{tconnor@carnegiescience.edu}

\author[0000-0002-7898-7664]{Thomas Connor}
\affil{The Observatories of the Carnegie Institution for Science, 813 Santa Barbara St., Pasadena, CA 91101, USA}

\author[0000-0003-4727-4327]{Daniel D. Kelson}
\affiliation{The Observatories of the Carnegie Institution for Science, 813 Santa Barbara St., Pasadena, CA 91101, USA}

\author[0000-0003-4218-3944]{Guillermo A. Blanc}
\affil{The Observatories of the Carnegie Institution for Science, 813 Santa Barbara St., Pasadena, CA 91101, USA}
\affil{Departamento de Astronom\'ia, Universidad de Chile, Camino del Observatorio 1515, Las Condes, Santiago, Chile}

\author[0000-0003-4432-5037]{Konstantina Boutsia}
\affiliation{Carnegie Observatories, Las Campanas Observatory, Casilla 601, La Serena, Chile}

\begin{abstract}

We present a catalog of spectroscopic redshifts for SPT-CLJ0615$-$5746, the most distant cluster in the Reionization Lensing Cluster Survey (RELICS). Using Nod \& Shuffle multi-slit observations with LDSS-3 on Magellan, we identify ${\sim}50$ cluster members and derive a cluster redshift of $z_c = 0.972$, with a velocity dispersion of $\sigma = 1235 \pm 170\ \textrm{km}\ \textrm{s}^{-1}$. We calculate a cluster mass using a $\sigma_{200}-M_{200}$ scaling relation of $M_{200} = (9.4 \pm 3.6) \times 10^{14}\ M_\odot$, in agreement with previous, independent mass measurements of this cluster. In addition, we examine the kinematic state of SPT-CLJ0615$-$5746, taking into consideration prior investigations of this system. With an elongated profile in lensing mass and X-ray emission, a non-Gaussian velocity dispersion that increases with clustercentric radius, and a brightest cluster galaxy not at rest with the bulk of the system, there are multiple cluster properties that, while not individually compelling, combine to paint a picture that SPT-CLJ0615$-$5746 is currently being assembled.
\end{abstract}


\section{Introduction} \label{sec:intro}
As the costs of building larger and more advanced telescopes swell \citep{2004SPIE.5489..563V}, gravitational lensing has emerged as an effective tool to maximize the returns of current telescopes and enable the study of fainter galaxies with already existing facilities. In particular, several large programs \citep{2012ApJS..199...25P, 2017ApJ...837...97L} on the \textit{Hubble Space Telescope} (HST) have surveyed clusters of galaxies, which are powerful cosmic lenses, with the aim of directly seeing galaxies in the epoch of reionization \citep[e.g.,][]{2014ApJ...792...76B, 2017ApJ...835..113L}. The utility of these lenses is based on the quality of the lensing model for the cluster, which therefore remains a key science product of these programs \citep{2014ApJ...795..163U, 2015MNRAS.452.1437J, 2018ApJ...859..159C}.

In addition to observations of highly-redshifted galaxies, these surveys have also produced fantastic imaging data for studies of the lensing clusters themselves. The scientific output of these programs includes studies of individual galaxies \citep[e.g.,][]{2012ApJ...756..159P, 2017ApJ...835..216D}, cluster properties \citep[e.g.,][]{2015ApJ...806....4M, 2018MNRAS.474.3009D}, and the suite of observed clusters \citep[e.g.,][]{2015ApJ...805..177D, 2017ApJ...848...37C}. When combined with other multi-wavelength observations, the rich data set of these surveys has allowed detailed analyses of the structure and formation of clusters \citep[e.g.,][]{2017ApJ...849...59B, 2018ApJ...861...71S}.

However, one major limiting factor for HST cluster studies is the need for spectroscopic followup. Redshift information is needed to not only study the clusters \citep[e.g.,][]{2019ApJ...875...16C}, but it is also required for obtaining the best possible lens models \citep{2017MNRAS.470.1809A, 2018MNRAS.473..663M, 2018ApJ...863...60R}. To that end, complementary spectroscopic surveys are needed in conjunction with these HST programs \citep{2014ApJS..211...21E, 2014Msngr.158...48R, 2015ApJ...812..114T}.

\begin{deluxetable*}{rrrrrrrr}
\tablecaption{Magellan Observing Log}
\tablewidth{0pt}
\tablehead{
\colhead{Mask ID} & \colhead{Instrument} & \colhead{Date\tablenotemark{a}} & \colhead{Dwell Time} & \colhead{Exposure Time} & \colhead{Seeing\tablenotemark{b}} & \colhead{$N_{Obs}$\tablenotemark{c}} & \colhead{$N_{z}$\tablenotemark{d}}\\
\colhead{} & \colhead{} & \colhead{(YYYY-MM-DD)} & \colhead{(s)} & \colhead{(s)} & \colhead{(arcsec)} & \colhead{} & \colhead{}}
\startdata
0 & LDSS-3 & 2017-11-30 & 90 & $2 \times 1800$ & 0.8 & 21 & 5 \\
1 & LDSS-3 & 2017-11-30 & 90 & $2 \times 1800$ & 1.0 & 23 & 5\\
2 & GISMO & 2018-01-13 & \nodata & $2 \times 2700$ & 0.8 & 34 & 11 \\
3 & LDSS-3 & 2018-01-14 & 45 & $2 \times 1800$ & 0.9 & 16 & 11\\
  &   & 2018-01-14 & 60 & $1 \times 1800$ & 1.0 & & \\
4 & LDSS-3 & 2018-01-14 & 60 & $3 \times 1800$ & 1.0 & 16 & 9\\
5 & LDSS-3 & 2018-01-14 & 60 & $3 \times 1800$ & 1.0 & 15 & 1\\
6 & LDSS-3 & 2018-01-14 & 60 & $1 \times 1800$ & 1.0 & 13 & 4\\
  &   & 2018-01-14 & 45 & $1 \times 1800$ & 1.1 & & \\
7 & LDSS-3 & 2019-01-13 & 45 & $4 \times 1800$  & 1.0 & 17 & 11\\
8 & LDSS-3 & 2019-01-13 & 45 & $4 \times 1800$ & 1.0 & 12 & 10\\
9 & LDSS-3 & 2019-01-13 & 45 & $2 \times 1800$ & 1.0 & 13 & 4\\
 &   & 2019-01-13 & 45 & $1 \times 1350$ & 1.0  & &\\
\enddata
\label{tab:ObsLogs}
\tablenotetext{a}{Date at start of the night}
\tablenotetext{b}{Worst seeing reported by the observer for those observations}
\tablenotetext{c}{Number of objects on mask}
\tablenotetext{d}{Number of objects with measured redshifts}
\end{deluxetable*}

One of the most recent large HST programs imaging clusters is the Reionization Lensing Cluster Survey \citep[RELICS,][]{2019arXiv190302002C}. Targeting 41 massive clusters, this survey has detected over 300 galaxies with photometric redshifts beyond $z > 6$ \citep{2017arXiv171008930S}. At the tail end of the redshift distribution for RELICS clusters is \longname. Independently discovered by the South Pole Telescope survey \citep{2011ApJ...738..139W} and \citet[][identified in that survey as PLCK G$266.6{-}27.3$]{2011A&A...536A..26P}, \longname\ (hereafter \clj) is a massive cluster ($M_{500} \sim 8 \times 10^{14}$; \citealt{2011A&A...536A..26P}) with a previously-measured redshift of $z\sim 0.97$ \citep{2011ApJ...738..139W}. Not only is \clj\ the second-most productive RELICS cluster for lensing $z > 6$ galaxies \citep{2017arXiv171008930S}, but \citet{2018ApJ...864L..22S} reported the detection of a $z \sim 10$ galaxy candidate strongly lensed by the cluster, which is potentially one of the most distant galaxies yet discovered.

\citet{2018ApJ...863..154P} previously studied \clj; they used the RELICS imaging to produce a strong lensing model for the cluster, which was based on three multiply imaged background galaxies and included contributions from cluster galaxies (as selected by the red sequence). This analysis found weak evidence to support a possible foreground concentration at $z \sim 0.4$, but lacked the spectroscopic followup needed to confirm the presence of such a structure. \citet{2018MNRAS.474.2635S} performed a weak lensing analysis of this cluster, and they reported that the cluster has an elongated or perturbed morphology. Additionally, \citet{2017A&A...598A..61B} reported that \clj\ has a high core-excised X-ray temperature: $T_X = 11.04 \pm 0.56\ {\rm keV}$. Both the cluster morphology and the high temperature are indicative of a major merger \citep[e.g.,][]{2001ApJ...561..621R,2018arXiv181207481O}. A redshift census of this cluster is necessary therefore not only to further the cluster science goals of RELICS, but also to refine the cluster lensing models and to provide insight into the kinematic state of the cluster.

\begin{figure*}[th]
\begin{center}
\includegraphics{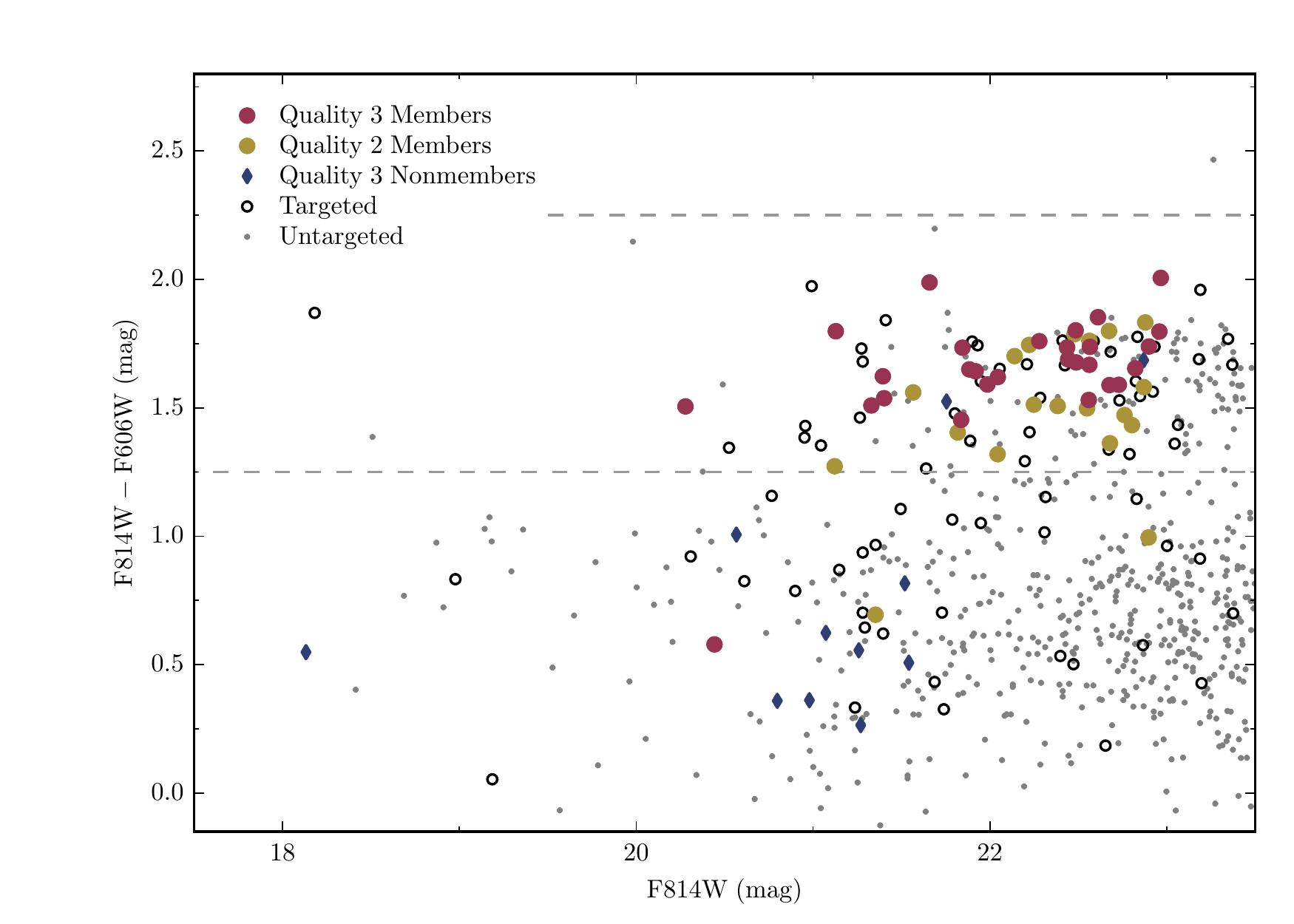}
\end{center}
\caption{Color-magnitude diagram for the field of \clj, using the photometry from the RELICS catalog. Galaxies are colored based on the results of our spectroscopic campaign. Cluster members with Quality 3 and 2 redshifts are marked by red and gold circles, respectively, while nonmembers with Quality 3 redshifts are indicated by blue diamonds. Galaxies that we observed but could not establish a redshift for are shown with open circles; note that galaxies were observed across different setups, and so galaxy magnitude should not be taken as a tracer of success. Due to the high redshift of \clj, the red sequence is poorly-defined in this filter set, but the next-reddest filter was observed using the WFC3/IR instrument and therefore has a much smaller field of view. The region of our red sequence color selection is marked by horizontal dashed lines.} \label{fig:CMD}
\end{figure*}

In this work, we describe a recent multi-object spectroscopic survey with the Magellan telescopes to determine redshifts for galaxies in the \clj\ field of view. We describe our observations in Section \ref{sec:obs}, and present an analysis of the kinematics of the cluster in Section \ref{sec:results}. Finally, we put this result in context with previous analyses of this cluster in Section \ref{sec:discussion}. Based on the analysis we present below, we adopt a cluster redshift of $z_c = 0.972$. Throughout this work, we adopt a flat $\Lambda$CDM cosmology with $\Omega_M = 0.3$ and ${\rm H}_0 = 70\ {\rm km}\ \textrm{s}^{-1}\ \textrm{Mpc}^{-1}$, so that at this redshift, $1\arcsec = 7.953\ \textrm{kpc}$. All magnitudes used in this work are AB, and all velocity dispersions have been corrected to account for broadening due to errors in redshifts following \citet{1980A&A....82..322D}.

\section{Observations} \label{sec:obs}

We observed the field of \clj\ across three separate observing runs using both Magellan 6.5m telescopes. A log of the observations is given in Table \ref{tab:ObsLogs}. To create a catalog of potential targets for observations, we used the photometric catalog assembled by the RELICS team\footnote{\url{https://archive.stsci.edu/prepds/relics/}}, using the galaxies detected in the combined Advanced Camera for Surveys (ACS) and Wide-Field Camera (WFC3) observations. We used the photometric redshifts from that catalog, derived using the code Bayesian Photometric Redshifts \citep[BPZ,][]{2000ApJ...536..571B,2004ApJS..150....1B,2006AJ....132..926C}. Slit masks were designed using \texttt{Maskgen}\footnote{\url{https://code.obs.carnegiescience.edu/maskgen}}. 

Our input catalogs were designed to maximize the number of cluster galaxies observed; we used the information we had to put the galaxies with the highest odds of being near $z=0.97$ on slits. While the exact methods to compute observing priority varied across our different runs and between instruments, the basic idea was the same: we gave the highest priorities to galaxies on the red sequence and to galaxies that had photometric redshifts near that of the assumed cluster redshift. As \texttt{Maskgen} uses a magnitude-like priority system, the base priority of each galaxy was its ${\rm F}814{\rm W}$\ magnitude, with a limiting magnitude of ${\rm F}814{\rm W} < 23.0$\ (except on Mask 2). Due to the large spread in observed colors, and since we had no confirmed members to define a model red sequence with, we defined the red sequence as existing in the color region of ${\rm F}606{\rm W} - {\rm F}814{\rm W} = 1.75 \pm 0.5\ {\rm mag}$ and ${\rm F}814{\rm W} - {\rm F}105{\rm W} = 1.025 \pm 0.175\ {\rm mag}$. As can be seen in Figure \ref{fig:CMD}, this color region is sufficient to capture red sequence galaxies. We also gave added priority to those galaxies with photometric redshifts near the cluster, defined with $z_B$, the BPZ most-likely redshift, being in the range $0.85 < z_B < 1.2$ (in this redshift range, 4000 \AA\ is within the wavelength range of our spectroscopic coverage). In total, 37 galaxies were observed on multiple masks.

Because of how we selected galaxies, there are some holes in our population that should be accounted for in any analysis more sophisticated than what we present here. To maximize the number of cluster galaxies we could observe using nod \& shuffle, we did not prioritize blue galaxies (and only found three such cluster galaxies; see Figure \ref{fig:CMD}), we did not sample out to the virial radius, and not every target was observed for the same duration. For this paper, the chief concern is that our sample is primarily composed of passive galaxies. However, previous work \citep[e.g.,][]{2017ApJ...837...88B} has shown that this should not bias our measurement of the overall cluster velocity distribution, at least within the precision we measure.
Below we describe the individual techniques used for the two instruments.

\begin{deluxetable*}{rllrrrrrrr}
\tablecaption{Redshift Catalog}
\tablewidth{0pt}
\tablehead{
\colhead{RELICS ID} &  \colhead{$\alpha_{2000}$} & \colhead{$\delta_{2000}$} & \colhead{z} & \colhead{Mask ID\tablenotemark{a}} & \colhead{Quality\tablenotemark{b}} & \colhead{${\rm F}606{\rm W}$\tablenotemark{c}} & \colhead{${\rm F}814{\rm W}$\tablenotemark{c}} & \colhead{${\rm F}105{\rm W}$\tablenotemark{c}} & \colhead{Emission\tablenotemark{d}}} 
\startdata
 3044 & 93.959124 & $-$57.813138 & 0.47715 & 2 & 3 & 21.696 & 21.072 & \nodata & 1\\
 3552 & 93.957319 & $-$57.808954 & 0.97692 & 4 & 3 & 24.476 & 22.822 & \nodata & 0\\
 3880 & 93.913769 & $-$57.807100 & 0.97087 & 8 & 3 & 24.319 & 22.729 & \nodata & 0\\
 3895 & 93.945245 & $-$57.807213 & 0.98590 & 3 & 3 & 24.638 & 22.899 & \nodata & 0\\
 4577 & 93.977897 & $-$57.805366 & 0.95168 & 4 & 2 & 22.395 & 21.122 & \nodata & 0\\
 5069 & 93.962891 & $-$57.801253 & 0.96553 & 3 & 3 & 24.090 & 22.559 & \nodata & 0\\
 5189 & 94.002923 & $-$57.800580 & 0.97059 & 8 & 2 & 24.451 & 22.870 & \nodata & 0\\
 5421 & 93.974158 & $-$57.799626 & 0.95619 & 3 & 3 & 22.839 & 21.330 & 20.787 & 0\\
 5633 & 93.977906 & $-$57.797646 & 0.98280 & 3 & 2 & 24.041 & 22.679 & 22.255 & 0\\
 5735 & 93.998374 & $-$57.797620 & 1.20441 & 2 & 3 & 22.335 & 21.519 & \nodata & 1\\
\enddata
\label{tab:Redshifts}
\tablenotetext{a}{Masks are detailed in Table \ref{tab:ObsLogs}}
\tablenotetext{b}{$3=\textrm{Good}$, $2=\textrm{Less certain}$, as described in the text.}
\tablenotetext{c}{AB magnitudes from the RELICS photometric catalog. Median uncertainties are $0.026$, $0.005$, and $0.009$ magnitudes for ${\rm F}606{\rm W}$, ${\rm F}814{\rm W}$, and ${\rm F}105{\rm W}$, respectively.}
\tablenotetext{d}{$1=\textrm{Emission}\ \textrm{lines}\ \textrm{observed}$, $0=\textrm{No}\ \textrm{emission}\ \textrm{lines}\ \textrm{observed}$.}

\tablecomments{Table \ref{tab:Redshifts} is published in its entirety in the machine-readable format. A portion is shown here for guidance regarding its form and content.}

\end{deluxetable*}

\subsection{LDSS-3 Nod and Shuffle}

The primary work of this survey was conducted with the Low Dispersion Survey Spectrograph 3 (LDSS-3) on the Magellan Clay telescope. Following a recent CCD upgrade \citep{2016ApJ...817..141S}, LDSS-3 has excellent red sensitivity beyond ${\sim}9000$ \AA\ with minimal fringing, making it an ideal instrument for investigating galaxies at redshift $z\sim1$. We conducted our observations using the Nod \& Shuffle mode \citep{2001PASP..113..197G}, which allows for exceptional background subtraction, even in the presence of numerous night sky lines. To maximize our ability to cover the relatively small WFC3 and ACS fields of view, we used a macro shuffle, whereby the charge is shuffled in units of one-third of the CCD. This shuffle scheme cuts our usable spectroscopic region into thirds: a storage region for the A nod position, the active region, and a storage region for the B nod position. Trisecting the ${\sim}8\farcm3$ LDSS-3 field of view brings the spatial coverage to approximately that of the WFC3/IR field. We used a small nod -- less than the length of the slit, as described below -- so that data were always being collected during an exposure for every galaxy being observed.

\begin{figure*}[t]
\includegraphics{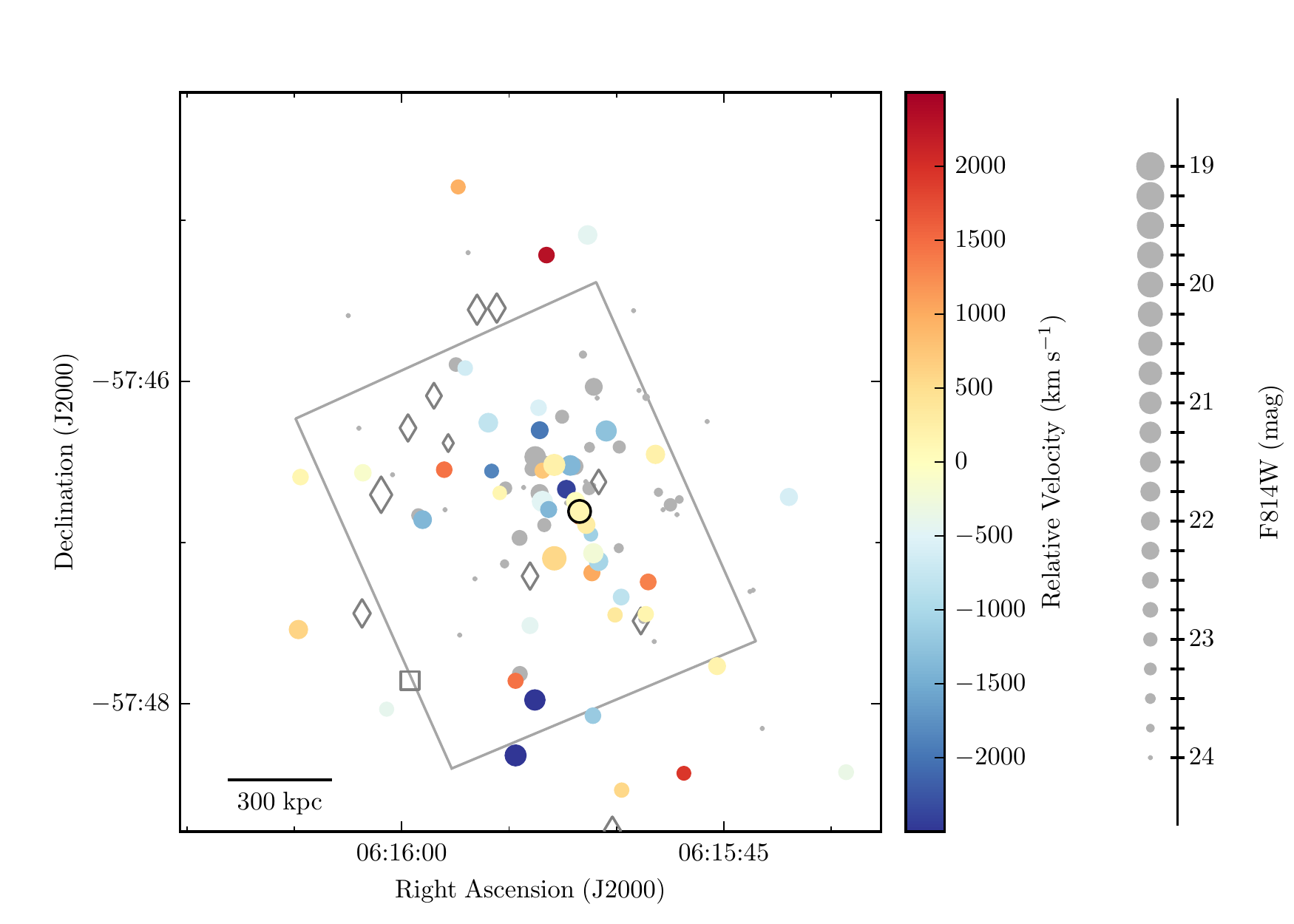}
\caption{Distribution of galaxies associated with \clj. Galaxies with spectroscopic redshifts are colored based on their velocity relative to the cluster. Those galaxies with absolute spectroscopic velocities greater than $3800\ {\rm km}\ {\rm s}^{-1}$ are marked by open diamonds (foreground) and squares (background). The BCG is indicated with a solid black border. Galaxies with photometric redshifts in the range $0.92 < z_B < 1.02$ and without spectroscopic information are marked by gray points. The sizes of each galaxy are proportional to their magnitude in the F814W filter (which has approximately the same wavelength coverage as our spectra), as shown by the key on the right of the figure. The WFC3/IR field-of-view is indicated by the faint gray square.} \label{fig:positions}
\end{figure*}

Masks were designed in \texttt{Maskgen} to cover the range 7200 \AA\ -- 9000 \AA\ to ensure coverage of the rest-frame 4000 \AA\ break; however, as there was no risk of slit collisions due to the narrow CCD, we observed all objects with only the OG590 order blocking filter. We observed each mask with the VPH-Red grism, which has spectral dispersion of $1.175\ \textrm{\AA}\ \textrm{pix}^{-1}$ and a resolving power of $R\approx 1350$. Alignment stars were selected from the RELICS photometry. All of our slits were $1\farcs0$ wide, but we adjusted the length after our 2017 November run. Initially, we used $4\farcs0$ long slits, with A and B positions separated by $1\farcs6$; upon testing, we learned that nodding in approximately integer pixel steps allowed for better image combination (the pixel scale at the LDSS-3 detector is $0\farcs189\ \textrm{pixel}^{-1}$). We therefore switched to $4\farcs5$ long pixels, with A and B positions separated by $1\farcs5$ for all masks observed in 2018 and beyond.

Our spectroscopic data were reduced following the method used by \citet{2018ApJ...867...25C}, which we briefly describe here. Global wavelength mappings were bootstrapped using Coherent Point Drift \citep[CPD,][]{2009arXiv0905.2635M} for the entirety of both shuffle positions. Then, slitlets were processed following the routines described in \citet{2000ApJ...531..159K} and \citet{2003PASP..115..688K}. Quartz-halogen lamp and HeNeAr spectral lamp images taken immediately following the observations were used for flat-fielding and wavelength calibration, respectively. Background subtraction was performed by subtracting the two nod positions from each other; stacked, rectified two-dimensional spectra were created from the multiple exposures, using local sigma clipping to reject cosmic rays. Once the spectra were reduced and extracted, we used the code described by \citet{2018AAS...23114942L} to calculate redshifts for each object using the spectral templates from the Sloan Digital Sky Survey \citep[SDSS,][]{2004AJ....128..502A}. 
\begin{figure*}[t]
\includegraphics{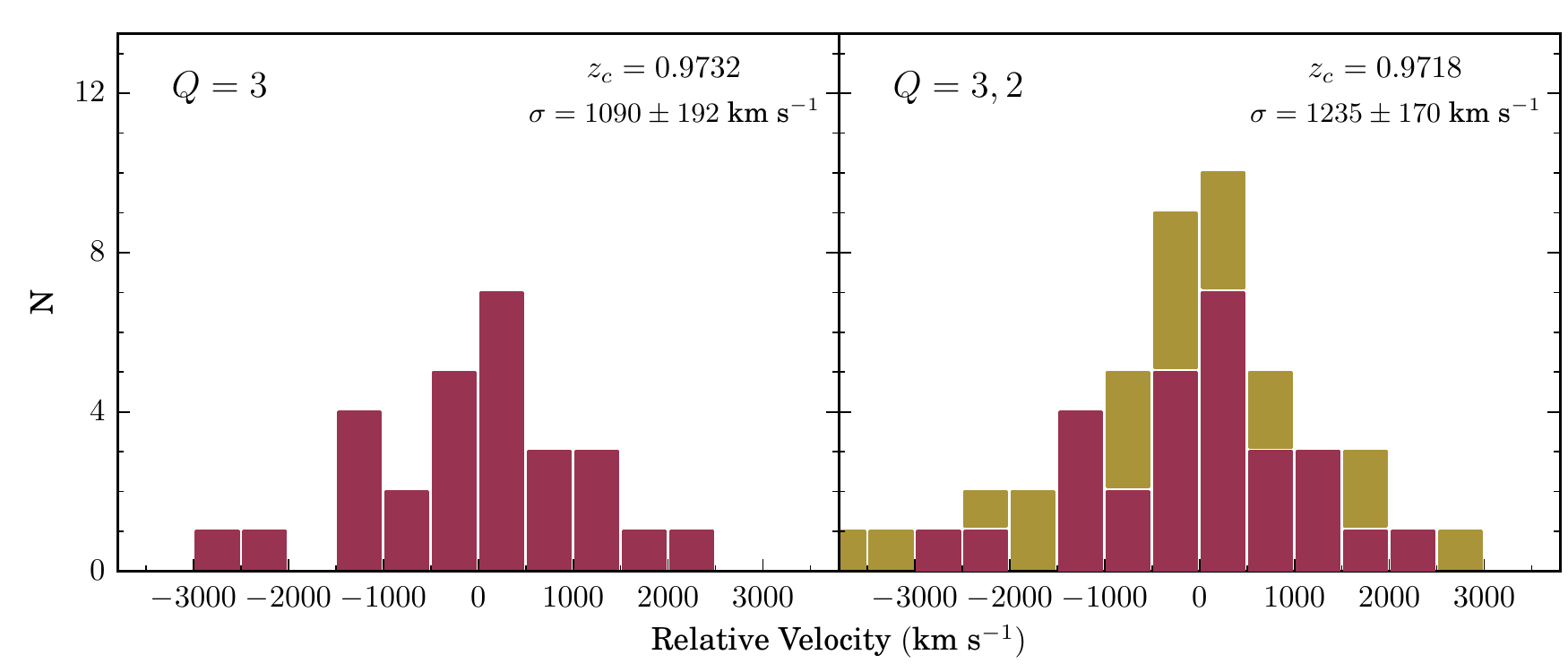}
\caption{Velocity distribution for our observed galaxies relative to the nominal cluster redshift for those galaxies of Quality 3 (\textbf{Left}) and of either Quality 3 and 2 (\textbf{Right}). Red and gold correspond to the number of galaxies with Quality 3 and 2 redshifts, respectively.} \label{fig:histogram}
\end{figure*}

\subsection{GISMO}
We also observed \clj\ with the Gladders Image-Slicing Multislit Option (GISMO) for the Inamori-Magellan Areal Camera and Spectrograph \citep[IMACS,][]{2011PASP..123..288D} on Magellan-Baade. With GISMO, we were able to observe a large number of objects in the small RELICS field of view, but without the ability to nod and shuffle and with a diminished throughput. Observations were conducted solely on 2018 Jan 13 and only for one mask. We used the {\it f/2} camera and the 300 l/mm Red grating. GISMO observations were made with the CTIO I filter, so that our observed spectral range was ${\sim}7000 - {\sim} 8750$ \AA. The spectral dispersion of this setup is $1.25\ \textrm{\AA}\ \textrm{pix}^{-1}$ with a resolving power of $R\approx 1280$

Our mask contained 34 objects with slits of width $1\farcs0$ and length $5\farcs0$. As with the LDSS-3 observations, data were reduced using the CPD code, though this time employing the B-spline sky modeling of \citet{2003PASP..115..688K} to do the sky subtraction, and redshifts were calculated with the tool of \citet{2018AAS...23114942L}. Details of the observation are given in Table \ref{tab:ObsLogs}.

\section{Results}
\label{sec:results}
In total, we observed 134 targets, of which 37 were observed on at least two separate masks. Of those 134 objects, we were able to measure redshifts for 58 galaxies and identified 6 M-class stars. These redshifts are presented in Table \ref{tab:Redshifts}. 5 galaxies had repeat observations with well-measured redshifts in both cases; their average offset (quantified as $1.4826\times\textrm{MAD}$, the median absolute deviation) was less than $1.4286 \times \widetilde{|\Delta z|} < 0.001$. 47 galaxies have velocities within $|\Delta(v)| \leq 3750\ \textrm{km}\ \textrm{s}^{-1}$ of the cluster (these galaxies are within the ${\sim}3\sigma$ region of the velocity dispersion, as discussed below), where velocities were measured using the formula
\begin{equation}
v = \frac{z - z_c}{1 + z_c} c,
\end{equation}
where $c$ is the speed of light. The distribution of these galaxies is shown in Figure \ref{fig:positions}.
\begin{figure*}[th]
\begin{center}
\includegraphics{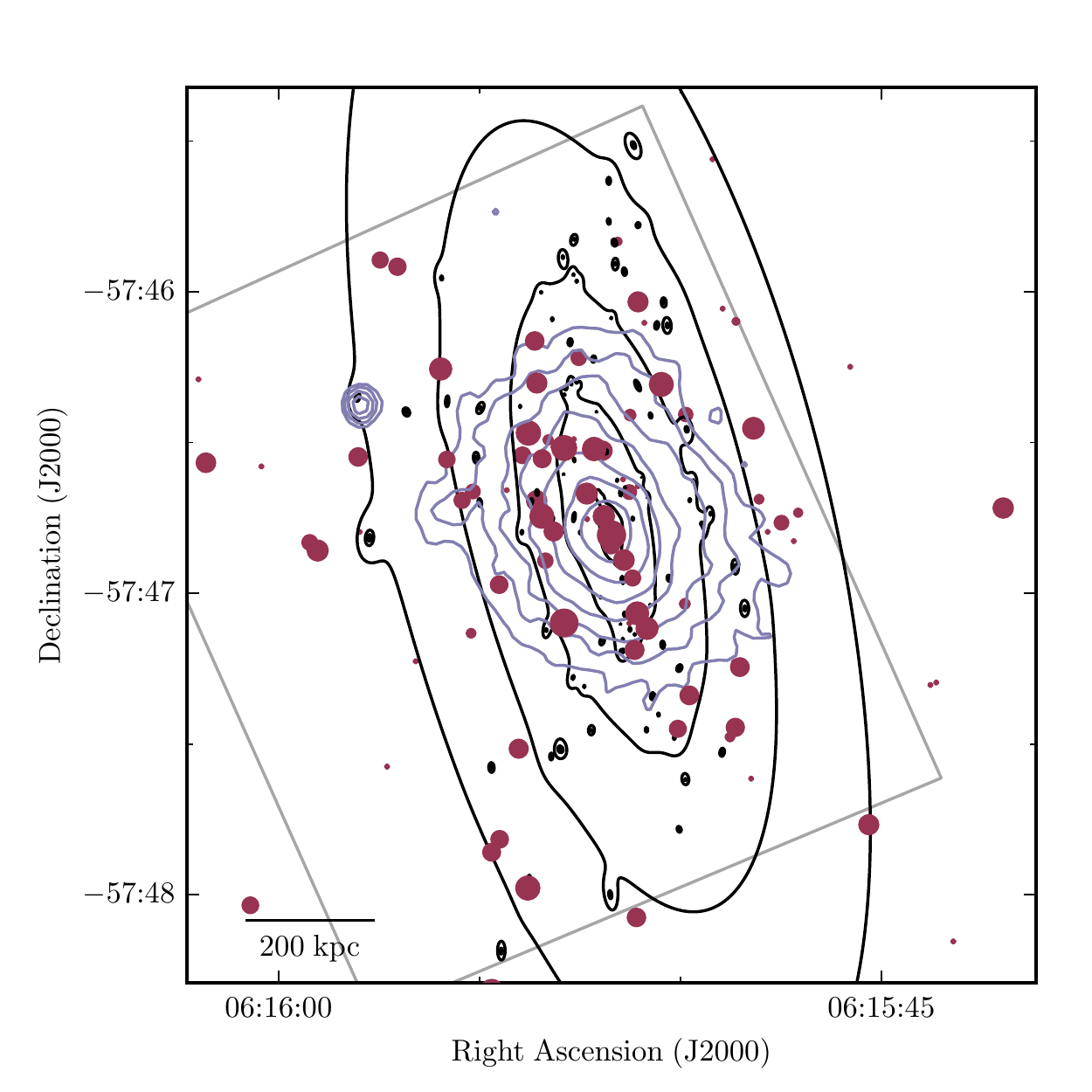}
\end{center}
\caption{The distribution of mass, galaxies, and gas around \clj. Spectroscopically-confirmed galaxies and those galaxies with photometric redshifts $0.92 < z_B < 1.02$ are marked with red points, with sizes given matching Figure \ref{fig:positions}. Black contours trace the mass surface density ($\kappa$, the convergence) reported by \citet{2018ApJ...863..154P}, with logarithmically-spaced contours. Smoothed $0.5 - 7.0$ keV emission from a ${\sim}225$ ks observation with {\it Chandra} \citep{2011cxo..prop.3451M} is shown with logarithmically-spaced blue contours. The image is centered on the BCG. For reference, the WFC3/IR field-of-view is indicated by the light gray square, as in Figure \ref{fig:positions}.} \label{fig:three_data}
\end{figure*}

We assigned a quality flag to all of our spectra to indicate the reliability of their measured redshifts, as given in Table \ref{tab:Redshifts}. While our flags are similar to those created by \citet{1995ApJ...455...60L} and used in many surveys since, we note that we lack the number of both overall measurements and repeat measurements to assign a flag based on inferred numerical probabilities; instead our flags are qualitative in nature. Quality 3 spectra have clearly identifiable emission line features or strong Calcium H and K absorption lines.

We also include galaxies with less secure redshifts as Quality 2 objects. While not by design, all of our Quality 2 sample have measured redshifts similar to that of \clj\ ($0.94 \lesssim z \lesssim 1.0$); since the cross-correlation redshift fitting has no knowledge of the expected redshift of the cluster, there is a low probability of a $z \sim 0.97$ measurement being reported for non-cluster galaxies. To minimize the effect of random chance on our Quality 2 redshifts, we varied the parameters of our redshift fitting (wavelength range used, redshift range evaluated) for all potential Quality 2 galaxies. For random alignments on noisy spectra, these variations were enough to move the measured redshift, but our reported Quality 2 redshifts were robust against this effect. Due to the gap between when we conducted our observations, we were able to target some Quality 2 objects for deeper observations. Of the six Quality 2 observations we were able to improve to Quality 3, the average offset was $1.4286 \times \widetilde{|\Delta z|} = 0.0036$. For the two Quality 2 observations for which we obtained another Quality 2 observation, there was still strong agreement, with an average offset of $1.4286 \times \widetilde{|\Delta z|} = 0.0030$. Based on this analysis, we adopt nominal redshift errors of $\sigma_z = 0.0010$ for Quality 3 redshifts and $\sigma_z = 0.0036$ for Quality 2 redshifts.
\begin{figure*}[th]
\includegraphics{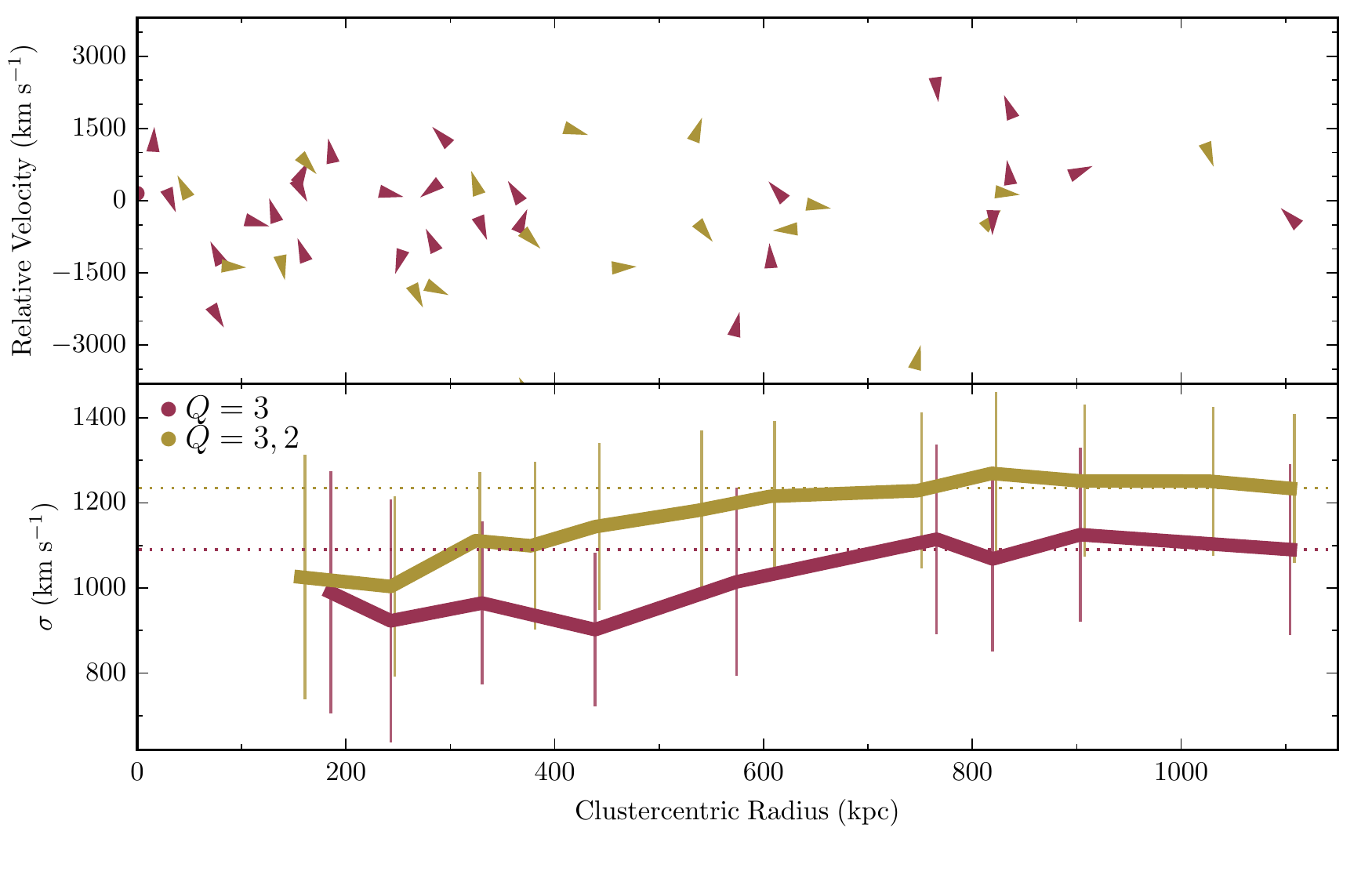}
\caption{\textbf{Top}: Measured line-of-sight velocity structure around \clj, colored by quality of the redshift (red symbols are Quality 3 and gold symbols are Quality 2). Galaxies are marked by arrows pointing in the direction on the sky from the galaxy to the BCG (which is shown by a circle); as a reference, the most distant galaxy shown is southwest of the BCG. We do not see any evidence for substructure in the galaxy phase-space distribution shown here. \textbf{Bottom}: Radial profile of the measured velocity dispersion, for the Quality 3 sample (red) and the full Quality 3 and 2 sample (gold). Errors are bootstrapped from the data. The adopted $\sigma$ values for both data sets are marked by dotted lines.} \label{fig:radial_vdisp}
\end{figure*}

To compute a cluster redshift, $z_c$, we use the biweight estimator of \citet{1990AJ....100...32B}. We do this first for only those galaxies with Quality 3 redshifts, and then for the full sample of Quality 3 and 2 objects. As there are no galaxies with measured velocities between $4000$ and $8000\ \textrm{km}\ \textrm{s}^{-1}$ of the cluster redshift, we select as cluster members all galaxies with relative velocities less than $4000\ \textrm{km}\ \textrm{s}^{-1}$. For the Quality 3 redshifts, we calculate $z_c = 0.9732$ and $\sigma = 1090 \pm 192\ \textrm{km}\ \textrm{s}^{-1}$; for Qualities 3 and 2, these values are $z_c = 0.9718$ and $\sigma = 1235 \pm 170\ \textrm{km}\ \textrm{s}^{-1}$. The velocity distribution of both sets is shown in Figure \ref{fig:histogram}. Uncertainties on the velocity dispersions are generated from smoothed bootstrap resampling of the input data, accounting for the nominal redshift errors. 

In order to compare our velocity dispersion with other works, we need to first standardize it to an aperture of $R_{200}$.
To our knowledge, no measurements of $R_{200}$ have been reported for \clj. However, by adopting the approximation $R_{500} \sim 0.7 R_{200}$ \citep[e.g.,][]{2009A&A...496..343E}, we convert previous measurements of $R_{500}\sim 1000\ \textrm{kpc}$ \citep{2017A&A...598A..61B, 2019ApJ...871...50B} to an estimate of $R_{200}\sim 1400\ \textrm{kpc}$. \citet{2016MNRAS.461..248S} reported on adjusting from velocity dispersions measured inside a given radius, $r_{max}$, to $R_{200}$. As we sample out to $r_{max} \sim 1000\ \textrm{kpc}$, our measured $\sigma_v$ should only differ from $\sigma_{200}$ by ${\sim} 2 \pm 4 \%$, so we adopt $\sigma_v \sim \sigma_{200}$. To account for the halo-to-halo variance, we also conservatively add an additional 5\% error to the uncertainty on $\sigma_{200}$.

We adopt the relation between $\sigma$ and $M_{200}$ of 
\begin{equation} \label{eqn:MfromSig}
\frac{M_{200}}{10^{15}\ M_\odot} = \frac{1}{h E(z)}\left( \frac{\sigma_{200}}{A_{1D}}\right)^{1/\alpha},
\end{equation}
where $\alpha$ and $A_{1D}$ are fitted parameters, h is assumed to be $h = 0.7$, and $E(z) = \left( (1 + z)^3 \Omega_M + \Omega_\Lambda \right)^{1/2}$. From simulations, \citet{2013MNRAS.430.2638M} report best-fits of $A_{1D} = 1177 \pm 4.2\ \textrm{km}\ \textrm{s}^{-1}$ and $\alpha = 0.364 \pm 0.0021$ for galaxy particles, with an expected ${\sim}15\%$ scatter on $\sigma$ around the fit, which we add in quadrature to our uncertainties. Using our values of $\sigma_{200}$, we find $M_{200}  = (6.7 \pm 3.3) \times 10^{14}\ M_\odot$ for Quality 3 objects and  $M_{200}  = (9.4 \pm 3.6) \times 10^{14}\ M_\odot$ for Quality 3 and 2 objects. These values -- particularly the latter -- are in the range expected from SZ measurements \citep{2015ApJS..216...27B}, weak lensing \citep{2018MNRAS.474.2635S}, strong lensing \citep{2018ApJ...863..154P}, and scaling from $Y_X$ \citep{2017A&A...598A..61B}. Below, we discuss how \clj\ shows signs of not being in virial equilibrium, although we expect that being out of equilibrium should not affect our mass estimate or its uncertainty. \citet{2018MNRAS.474.3746A} found that the intrinsic scatter in the $\sigma_{200} -M_{200}$ relation for simulated galaxies does not have a statistically significant difference between relaxed and non-relaxed clusters. And, as \citet{2013MNRAS.430.2638M} calibrated this relation with all clusters, not just relaxed ones, we expect that any deviation caused by the kinematic state of the cluster is already captured by the scatter, particularly considering how large the uncertainties in the mass already are.

\section{Discussion}
\label{sec:discussion}

In order to contextualize \clj\ among other galaxy clusters, we need to answer a fundamental question: is \clj\ an ongoing merger? One of the best analogs of this cluster -- in mass, X-ray temperature, and lensing ability -- is MACS J$0717.5{+}3745$, itself a prominent merging cluster \citep[e.g.,][]{2013ApJ...777...43M}. One way to quantify the state of the cluster is through the Z-score \citep{1991ApJ...383...72G}, which measures the offset of the brightest cluster galaxy (BCG) from the rest of the cluster's velocity dispersion. Following the convention that a velocity offset is significant when the 90\% confidence intervals do not bracket 0 \citep[e.g.,][]{1994ApJ...422..480B}, we do not find evidence for a disturbed core in the Quality 3 data ($Z = 0.14^{+0.40}_{-0.25}$, 90\% confidence) but do with the Quality 3 and 2 data ($Z = 0.29^{+0.24}_{-0.26}$, 90\% confidence). An important note about this result is that we find a different BCG redshift than \citet{2011ApJ...738..139W}, who report $z_{BCG} = 0.972$ (whereas we find $z_{BCG} = 0.9742$). As that work only reports that their redshift determination was based on a longslit observation with IMACS on the Magellan-Baade telescope, and does not include information about observational setup, exposure times, or observing conditions, we cannot attempt to determine why our results are in disagreement.

X-ray observations of \clj\ show that the cluster is hot; \citet{2017A&A...598A..61B} report a temperature within $[0.15 - 0.75] R_{500}$ of $T_X = 11.04 \pm 0.56\ {\rm keV}$. \citet{2019ApJ...871...50B} also reported X-ray results for \clj, finding a core-included X-ray temperature of $T_X = 14.16_{-1.32}^{+2.04}$ keV and a core-excised temperature of $T_X = 12.50_{-1.99}^{+1.60}$ keV. While \clj\ is very massive, these temperatures imply an even more massive cluster, if it was in virial equilibrium. Adopting the scaling relation of \citet{2016MNRAS.463..413W}, we would expect a velocity dispersion of $\sigma \sim 2000\ \textrm{km}\ \textrm{s}^{-1}$ for a cluster of this temperature. Ongoing mergers have been seen to boost X-ray temperatures in simulations \citep[e.g.,][]{2001ApJ...561..621R, 2002MNRAS.329..675R}, and this effect requires a relatively recent merger to have occurred or to be occurring \citep{2002ApJ...577..579R}. While \citet{2017A&A...598A..61B} found that \clj\ is morphologically-relaxed in X-rays as measured by the centroid shift parameter $\langle w\rangle$, this result was based on a shallow {\it XMM-Newton} observation ($\sim 10\ \textrm{ks}$) and a not statistically-significant measurement with \textit{Chandra} ($\langle w\rangle = 0.0094 \pm 0.0011$, where disturbed clusters are those with $\langle w\rangle > 0.01$).

The mass distribution of \clj, as measured by lensing, appears to be elongated. In the two best strong-lensing models reported by \citet{2018ApJ...863..154P}, the ellipticity $\epsilon$ of the mass distribution is $\epsilon = 0.55^{+0.01}_{-0.05}$ or $\epsilon = 0.71^{+0.10}_{-0.01}$, oriented roughly ${\sim}30^\circ$ E of N. In their weak-lensing analysis of 13 high redshift clusters, \citet{2018MNRAS.474.2635S} identified \clj\ as one of five clusters with elongated or disturbed morphologies. While no quantification of this elongation is given, visual inspection shows that the elongation in the weak lensing map shown by \citet{2018MNRAS.474.2635S} is roughly oriented in the same direction as reported by \citet{2018ApJ...863..154P}. There is superficial evidence of this orientation and elongation also appearing in the \textit{Chandra} observations reported on by \citet{2017A&A...598A..61B}, but this, too, has not been quantified. Nevertheless, through visual inspection of the X-ray emission, we derive rough estimates for this morphology: the ellipticity is $\epsilon \approx 0.60 - 0.75$ oriented ${\sim}20^\circ$ to ${\sim}30^\circ$ E of N. We show the mass distribution from strong lensing, the X-ray emission, and the positions of cluster galaxies in Figure \ref{fig:three_data}.

One key question is whether our data are consistent with an underlying Gaussian distribution. To test for normality, we employ the Anderson-Darling test \citep{anderson1952}; previous analysis by \citet{2009ApJ...702.1199H} identified this test as the best statistical tool for analyzing galaxy system dynamics with a limited population of measured velocities, as it is reliable for samples of $n \geq 5$. For both samples we consider in this work, the squared Anderson-Darling test statistic, $A^2$, exceeds that of the $p=0.01$ significance levels, implying that the velocity distribution of cluster galaxies is non-Gaussian. In an analysis of different relaxation indicators for high-mass clusters, \citet{2018MNRAS.475.4704R} found that a high value of $A^2$ correlates with X-ray asymmetry but only marginally with $\langle w\rangle$, which is what we see here. 

Another way of quantifying the kinematic state of \clj\ with our data is to see how the measured velocity dispersion changes with clustercentric radius. Using the same techniques to estimate $\sigma$ as before, and with uncertainties taken from bootstrap re-sampling, we calculated velocity dispersions for increasing radii, as shown in Figure \ref{fig:radial_vdisp}. There is a trend of increasing dispersion with radius, although it is not larger than the statistical uncertainties. Nevertheless, our velocity dispersion profile follows what is expected by \citet{2009ApJ...702.1199H} for a non-Gaussian galaxy dispersion, and looks more like what is expected of a merging cluster than a non-merging cluster, as described by \citet{2018MNRAS.481.1507B}.

None of these diagnostics are, by themselves, conclusive of \clj\ being in the process of undergoing a merger. However, when taken as a whole, there is a compelling amount of evidence for the cluster being dynamically disturbed. Further spectroscopic study to obtain more cluster redshifts, a deeper X-ray observation with {\it XMM-Newton} to measure the extended gaseous structure, and improved lensing models will all contribute to clarifying the kinematic state of \clj. 

\acknowledgments
{\small This paper includes data gathered with the 6.5 meter Magellan Telescopes located at Las Campanas Observatory, Chile. This work is based on observations taken by the RELICS Treasury Program (GO 14096) with the NASA/ESA HST, which is operated by the Association of Universities for Research in Astronomy, Inc., under NASA contract NAS5-26555. The authors thank Katey Alatalo for donating unused time on Clay for further observations.}

\vspace{5mm}
\facilities{Magellan:Clay (LDSS-3), Magellan:Baade (IMACS:GISMO), HST}
\software{CosmoCalc \citep{2006PASP..118.1711W}, CarPy (\citealt{2000ApJ...531..159K}; \citealt{2003PASP..115..688K})}


\end{document}